\documentclass[preprint,11pt]{JHEP3}
\JHEPspecialurl{http://jhep.sissa.it/JOURNAL/JHEP3.tar.gz}

\usepackage{epsfig,multicol,amsmath}

\usepackage{bm}  % for the command \bm (boldface greek letters)
\usepackage{amsmath}     % defines the command \dfrac (fractions in displaystyle)
\usepackage{epsfig,epsf}
\usepackage{graphicx}
\usepackage{rotating}
\usepackage{cite}
\def\beq{\begin{equation}}
\def\eeq{\end{equation}}
\def\bea{\begin{eqnarray}}
\def\eea{\end{eqnarray}}

\def\eq#1{{Eq.~(\ref{#1})}}
\def\fig#1{{Fig.~\ref{#1}}}
\newcommand{\bas}{\bar{\alpha}_S}
\newcommand{\as}{\alpha_S}

\newcommand{\Lb}{\left(}
\newcommand{\Rb}{\right)}
\newcommand{\h}{\frac{1}{2}}
\parskip=\itemsep               %?

\newcommand{\nn}{\nonumber}

\newcommand{\ph}{\varphi}

\newcommand{\om}{\omega}

\def\pom{{I\!\!P}}
\def\reg{{I\!\!R}}
\begin{boldmath}
\title{BFKL  equation with running  QCD coupling and HERA data}
\end{boldmath}
\author{\Large 
Eugene\, Levin${}^{a, b}$ \thanks{Email: leving@post.tau.ac.il, eugeny.levin@usm.cl.}\,\,\,\,and\,\,Irina \,\,Potashnikova${}^{b}$ \thanks{Email: irina.potashnikova@usm.cl}\\
${}^a$ \, Department of Particle Physics, School of Physics and Astronomy,
Tel Aviv University, Tel Aviv, 69978, Israel\\
${}^b$\, Departamento de F\'\i sica,
Universidad T$\acute{e}$cnica Federico Santa Mar\'\i a   and
Centro Cient\'\i fico-Tecnol$\acute{o}$gico de Valpara\'\i so,
Casilla 110-V,  Valparaiso, Chile\\
}

\abstract
{ In this paper we developed approach based on the BFKL evolution in $\ln\Lb Q^2\Rb$. We show that the simplest diffusion approximation with running QCD coupling is able to describe the HERA experimental data on the
       deep inelastic structure function with good $\chi^2/d.o.f. \approx 1.3$.   From our description of the  experimental data we learned several lessons; (i)  the non-perturbative physics at long distances started to show up at $Q^2 = 0.25\,GeV^2$; (ii)  the scattering amplitude at  $Q^2 = 0.25\,GeV^2$ cannot be written as sum of soft Pomeron and the secondary Reggeon but the Pomeron interactions should be taken into account; (iii)  the Pomeron interactions can be reduced to the enhanced diagrams and, therefore, we do not see any needs for the shadowing corrections at HERA energies; and (iv) we demonstrated that the shadowing correction could be sizable at higher than HERA energies without any contradiction with our initial conditions.
          
  }

\keywords{Color Glass Condensate, gluon saturation, BFKL Pomeron, calculus,  non-linear evolution, geometric scaling behavior
}
\dedicated{PACS: 12.38-t, 12.38.Cy,1 2.38.Lg, 13.60.Hd, 24.85.+p, 25.30.Hm}
\preprint{TAUP 2972/13 \\
{\tt }\\
\today}

\begin{document}
%%%%%%%%%%%%%%%%%%%%%%%%%%%%%%%%%%%%%%%%%%%%%%%%%%%%%%%%%%%%%%%
\section{ Introduction}
%%%%%%%%%%%%%%%%%%%%%%%%%%%%%%%%%%%%%%%%%%%%%%%%%%%%%%%%%%%%%%%

High  energy (low $x$) deep inelastic scattering (DIS) probes the gluon density in the hadron.
Its energy evolution is determined by the BFKL equation\cite{BFKL,RevLI} which sums the leading log terms of the order of $\Lb \as \ln(1/x)\Rb^n$. During the past two decades different facets of the BFKL dynamics have been discussed in the  number of papers (see Res.\cite{RevLI,REV} for reviews).
In our opinion, such close attention to the BFKL dynamics  is rooted in two causes. First,  the increase of the gluons density at high energy ($\propto (1/x)^\lambda$)  has been observed experimentally at HERA\cite{HERA}; and we need to take into account the BFKL dynamics  to understand this increase. In other words, we  can view on the BFKL dynamics as the evolution\cite{GLR} of the gluon density at low Bjorken $x$ in DIS. However, the BFKL equation gives an inspiration or might be even the educated guess for the non-perturbative origin of the soft Pomeron contribution or,  in general,  it can create ideas about the high energy asymptotical behaviour of the scattering amplitude in the confinement region. We have even indication that BFKL equation generates the infinite number of Pomerons (Regge poles) for the running QCD coupling (see Refs. \cite{RevLI,GLR,LEV}). 

Recently, in the interesting papers (see Refs.\cite{KLR1,KLR2,KLRW}) the consistent approach, based on the point of view that the BFKL equation is the theory of the reggeons, has been developed and applied to description of the HERA data on DIS.  The successful representation of the data undermine the widespread prejudice that the BFKL evolution is not needed for a description of the HERA data (see Refs.\cite{CTEQ,MRST,ALE,HERAPDF}), In this paper we are going to hammer the last nail in the coffin of this prejudice showing that the  good  fit of HERA data is naturally appeared in the evolution equation approach to the BFKL dynamics.

%%%%%%%%%%%%%%%%%%%%%%%%%%%%%%%%%%%%%%%%%%%%%%%%%%%%%%%%%%%%%%%

\begin{boldmath}
\section{BFKL equation with running $\as$ as the evolution equation}
\end{boldmath}
%%%%%%%%%%%%%%%%%%%%%%%%%%%%%%%%%%%%%%%%%%%%%%%%%%%%%%%%%%%%%%%
\subsection{The equation}
%%%%%%%%%%%%%%%%%%%%%%%%%%%%%%%%%%%%%%%%%%%%%%%%%%%%%%%%%%%%%%%
The NLO BFKL equation can be written in the form (see \cite{KLR1,KLR2,FADLI})
\beq \label{RAS1}
\frac{\partial N\Lb k_\bot, Y\Rb}{\partial Y}\,\,=\,\,\bas\Lb k_\bot\Rb \int d^2 k'_\bot K_{LO}\Lb k_\bot, k'_\bot\Rb N\Lb k'_\bot, Y\Rb\,\,+\,\,\bas^2\Lb k_\bot\Rb \int d^2 k'_\bot K_{NLO}\Lb k_\bot, k'_\bot\Rb N\Lb k'_\bot, Y\Rb
\eeq
where 
\beq \label{RAS2}
N\Lb k_\bot, Y\Rb\,\,=\,\,\frac{1}{\sqrt{\bas\Lb k_\bot\Rb}}\,\int d^2 x\, e^{i \vec{k}_\bot\cdot \vec{x}}\,\int d^2 b\,\frac{N\Lb x, b; Y\Rb}{x^2}
\eeq
with $N\Lb r, b:Y\Rb$ being the imaginary part of the  scattering amplitude of the dipole with size $x$.
\beq \label{RAS3}
\bas\Lb k_\bot\Rb\,\,=\,\,\Lb N_c/\pi\Rb \as\Lb k_\bot \Rb\,\,=\,\,\frac{1}{b \ln\Lb k^2_\bot/\Lambda^2_{QCD}\Rb}
\eeq

and 
\beq \label{RAS4}
K_{LO}\Lb k_\bot, k'_\bot\Rb\,\,=\,\,\frac{1}{\Lb\vec{k}_\bot \,-\,\vec{k}'_\bot\Rb^2}\,\,-\,\,\frac{k^2_\bot}{\Lb\vec{k}_\bot \,-\,\vec{k}'_\bot\Rb^2 \Lb \Lb\vec{k}_\bot \,-\,\vec{k}'_\bot\Rb^2 + k'^2_\bot\Rb}\,\delta^{(2)}\Lb \vec{k}_\bot - \vec{k}'_\bot\Rb
\eeq
while $K_{NLO}\Lb k_\bot, k'_\bot\Rb$ is written in Ref.\cite{FADLI}.

One can see that in \eq{RAS1} we do not use the triumvirate structure\cite{TRI} of the LO BFKL for running $\bas$ which looks as follows:
\beq \label{RAS5}
\frac{\partial N\Lb k_\bot, Y\Rb}{\partial Y}\,\,=\,\, \int d^2 k'_\bot \Lb \frac{\bas\Lb k'_\bot\Rb \bas\Lb \vec{k}_\bot - \vec{k}'_\bot\Rb}{\bas\Lb k_\bot\Rb}\Rb\,K_{LO}\Lb k_\bot, k'_\bot\Rb N\Lb k'_\bot, Y\Rb\,\,
\eeq
The advantage of this expression that it preserves the bootstrap equations for the reggeized gluon that has been proven in the NLO BFKL approach\cite{FAFI}. On the other hand \eq{RAS5} takes into account part of the NLO corrections of \eq{RAS1} which are not the largest contribution to $K_{NLO}$ in \eq{RAS1}.
Since the main goal of this paper to clarify some rather qualitative features of the BFKL dynamics with running QCD coupling we feel it is reasonable to use the LO contribution to the simple equation (see \eq{RAS1}) following the example of Refs.\cite{KLR1,KLR2,KLRW}.

Finally, in this paper we are going to discuss the following equation:
\beq \label{RASF}
\frac{\partial N\Lb k_\bot, Y\Rb}{\partial Y}\,\,=\,\,\bas\Lb k_\bot\Rb \int d^2 k'_\bot K_{LO}\Lb k_\bot, k'_\bot\Rb N\Lb k'_\bot, Y\Rb
\eeq
with  $\bas\Lb k_\bot\Rb$ and $K_{LO}$ are given by \eq{RAS3} and \eq{RAS4}, respectively.

%%%%%%%%%%%%%%%%%%%%%%%%%%%%%%%%%%%%%%%%%%%%%%%%%%%%%%%%%%%%%%%

\subsection{Green function and  the set of Pomerons}

%%%%%%%%%%%%%%%%%%%%%%%%%%%%%%%%%%%%%%%%%%%%%%%%%%%%%%%%%%%%%%%

 In our approach treating the BFKL equation as evolution in $k_\bot$ we need to find a Green function  ($G_(Y, r )$)  which satisfies the following initial condition:
 \beq \label{GFIC} 
G_(Y - Y_0 , r = r_0)\,\,=\,\,\delta(Y \,-\,Y_0) \,\,\,\,\,\mbox{where}\,\,\,\,r\,\,\equiv\,\,\ln\Lb k^2_\bot/\Lambda^2_{QCD}\Rb
\eeq
  Using this function we can find the solution to the BFKL evolution equation ($N\Lb r, Y\Rb$) with given initial gluon distribution $N_{in}\Lb Y_0, r=r_0\Rb$
 \beq \label{SOLR}
N_{fin}(Y,k_T)\,\,=\,\,N_{fin}\Lb Y, r\Rb\,\,=\,\,\int \,d Y_0 \,G(Y - Y_0, r) \,\,N_{in}(Y_0, r = r_0)
\eeq 
In other word, \eq{SOLR} is a realization of the evolution in $r$.

We use the Mellin transform to find $G\Lb r, Y - Y_0\Rb$ in the for

\bea 
G\Lb Y - Y_0, r\Rb &=& \,\,\int^{a \,+\,i\infty}_{a - i\infty}\,\frac{d \om}{2 \pi
i}\,\,G\Lb \om, r\Rb \,e^{ \om \Lb  Y  -  Y_0\Rb}\label{mellin1}\\
G\Lb\omega, r \Rb\,&=&
\,\,\int^{a \,+\,i\infty}_{a - i\infty}\,\frac{d f}{2 \pi
i}\,\,g( \om,f)\,\,
\ph_f(r)\,\,=\,\, \,\int^{a \,+\,i\infty}_{a - i\infty}\,\frac{d f}{2 \pi
i}\,g(\om) \,e^{
\,\frac{1}{b\om}\,\int^{f_0}_{f}\,\chi (f') d f'\,\,+\,\,r\,f}\label{mellin2}
\eea
where $\chi\Lb f \Rb$ is the Mellin transform of the $K_{LO}$. Solution of \eq{mellin2} was firstly written in Ref.\cite{GLR} and has been discussed in details (see Refs.\cite{KLR1,KLR2 } and references therein).

In this paper we
 will proceed with the diffusion   approximation for $\chi\Lb f \Rb$ for the sake of simplicity. A generalization is simple and straightforward.
Therefore
\beq \label{CHI}
\chi\Lb f \Rb\,\,=\,\,\chi_0\,\,+\,\,D_0 \,(f - \h)^2 \,\,\,\,\,\,\mbox{with}\,\,\,\,\,\,\chi_0 \,=\,4 \ln2 = 2.772 \,\,\,\mbox{and}\,\,\,\,D_0\,=\,14 \zeta\Lb 3 \Rb \,=\,16.828
\eeq

The general solution is
\beq \label{RAS6}
G\Lb Y - Y_0, r \Rb\,\,=\,\,\int^{a + i\infty}_{a - i \infty} \,\frac{d \om}{
2 \,\pi\,i}\,\int^{f_0 + i \infty}_{f_0 - i \infty}\,\frac{d f}{2\,\pi\,i}\,
\tilde g(\om)\,e^{ \om \,(Y - Y_0) \,+\,f\,r
\,-\,\frac{\Lb\,\chi_0 f \,+\frac{D_0}{3}\,f^3\,\Rb}{b\,\om}}\,\,
\eeq

Denoting

\beq \label{RAS7}
A\Lb \om, r\Rb\,\,=\,\,\int^{f_0 + i \infty}_{f_0 - i \infty}\,\frac{d f}{2\,\pi\,i}\,
\,e^{ \,f\,r
\,-\,\frac{\Lb\,\chi_0 \,f \,+\,\frac{D_0}{3}\,f^3\,\Rb}{b\Lb r \Rb \,\om}}
\eeq

one can see that Green's function which satisfies \eq{GFIC} is equal to

\beq \label{GF}
G\Lb y, r\Rb \,\,\,=\,\,\,\int^{a + i\infty}_{a - i \infty} \,\frac{d \om}{
2 \,\pi\,i}\,e^{\om (y - y_0)}\,\frac{A\Lb \om, r\Rb}{A\Lb \om, r_0\Rb}
\eeq

For our simplified BFKL kernel

\beq \label{RAS8}
A\Lb \om, r\Rb\,\,=\,\,\Big( \frac{b\,\om}{D_0}\Big)^{1/3}\,Ai\Lb \Big( r \,\,-\,\,\frac{\chi_0}{ b \om}\Big) \Big( \frac{b\,\om}{D_0}\Big)^{1/3}\Rb
\eeq

\begin{figure}
  \leavevmode
      \includegraphics[width=14cm,height=6cm]{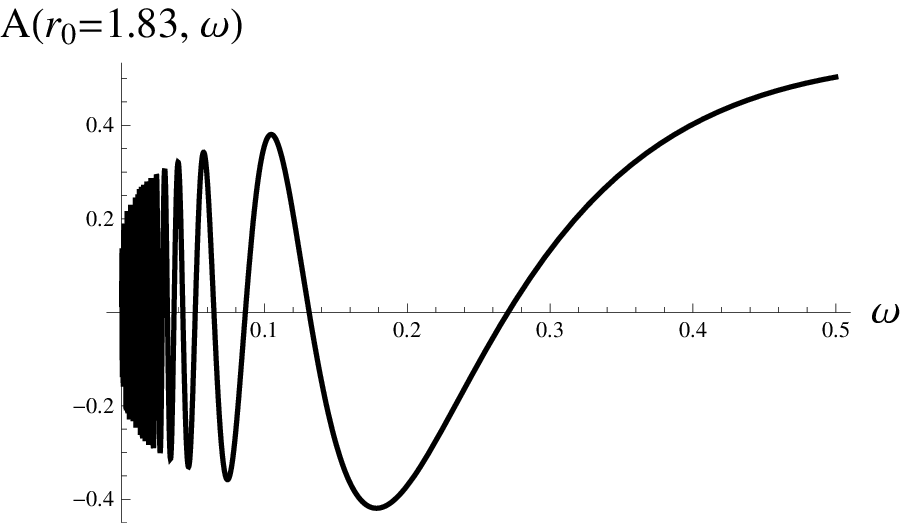}
      \caption{Function $A\Lb \om, r = r_0\Rb$ versus $\om$.  $r_0 = 1.83$}
      \label{1}
\end{figure}

One can see that this solution has a 
discrete spectrum\cite{LEV} of states that are  determined by the zeros of $A\Lb \om, r_0\Rb$ or by the roots of the following equation
\beq\label{RASS}
Ai\Lb \Big( r_0 \,\,-\,\,\frac{\chi_0}{ b  \om}\Big) \Big( \frac{b\,\om}{D_0}\Big)^{1/3}\Rb\,\,=\,\,0
\eeq
 In \fig{1} it is plotted function $A\Lb \om, r=r_0\Rb$ versus $\om$. One can see that we have the set of zeros which condenses to zero.
 %%%%%%%%%%%%%%%%%%%%%%%%%%%%%%%%%%%%%%%%%%%%%%%%%%%%%
 %%%%%%%%%%%%%%%%%%%%%%%%%%%%%%%%%%%%%%%%%%%%
  \begin{figure}
  \leavevmode
  \begin{tabular}{c c c}
      \includegraphics[width=6cm]{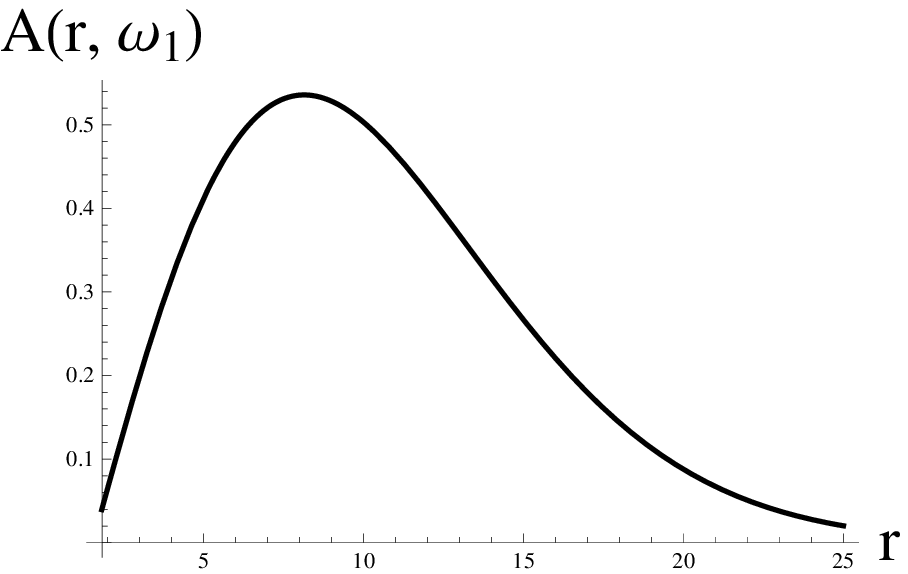}& \includegraphics[width=6cm]{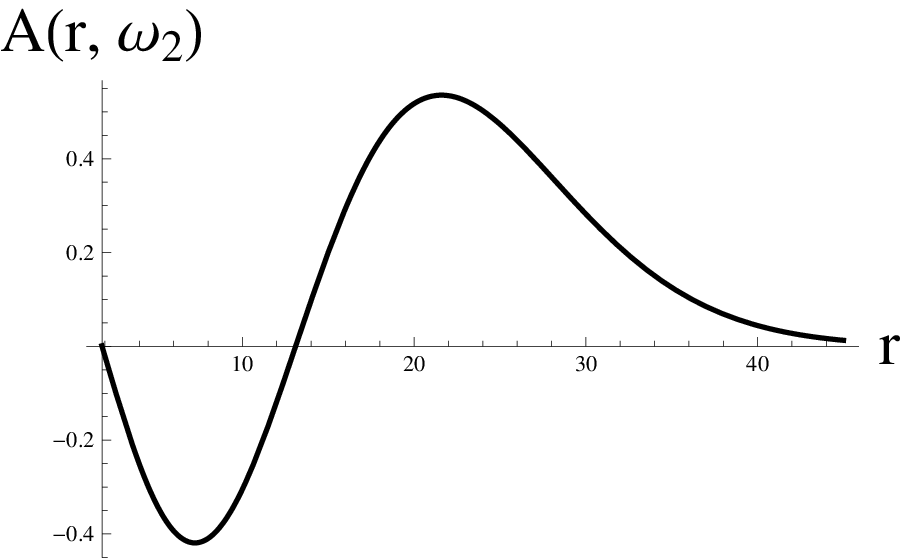}&  \includegraphics[width=6cm]{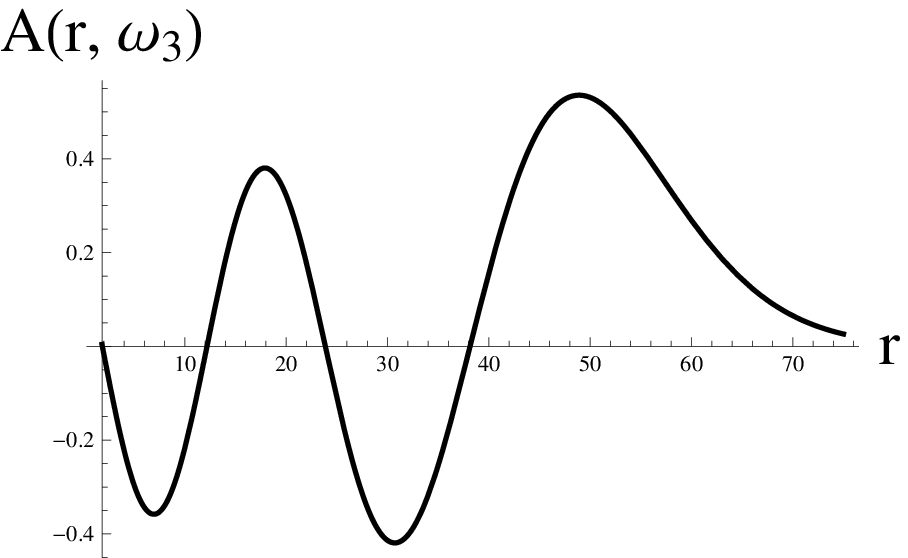} \\
   \fig{2}-a & \fig{2}-b &\fig{2}-c\\
   \end{tabular}
      \caption{Function $A\Lb \om_n, r \Rb$ versus $r$.  $r_0 = 1.83$ and $\om_1 = 0.27$,$\om_2 = 0.1312$ and $\om_3 = 0.0646$.}
      \label{2}
\end{figure}  

%%%%%%%%%%%%%%%%%%%%%%%%%%%%%%%%%%%%%%%%%%%%
  Airy functions have   zeros only at the negative values
of the argument,  and  their position   can be found with good
accuracy from the simple equation:
\beq \label{EQZE}
z\,\,=\,\,-  \, (\,\frac{3\pi n}{2} \,-\,\frac{3 \pi}{8}\,)^{\frac{2}{3}}
\,\,,
\eeq
Using \eq{EQZE} we can find the spectrum of the BFKL equation analytically, solving the equation
\beq \label{POLE}
\Lb \Big( r_0 \,\,-\,\,\frac{\chi_0}{ b \om_n}\Big) \Big( \frac{b\,\om_n}{D_0}\Big)^{1/3}\Rb
\,\,=\,\,-  \, (\,\frac{3\pi n}{2} \,-\,\frac{3 \pi}{8}\,)^{\frac{2}{3}}
\eeq

At large $n$ we have a  solution
\beq \label{POLE1}
\om_n\,\,=\,\,\frac{2 }{ 3 \pi b n}\,\Big( \frac{\chi_0}{ D_0^{1/3}}\Big)^{3/2}
\eeq

\begin{figure}
  \leavevmode
      \includegraphics[width=14cm,height=8cm]{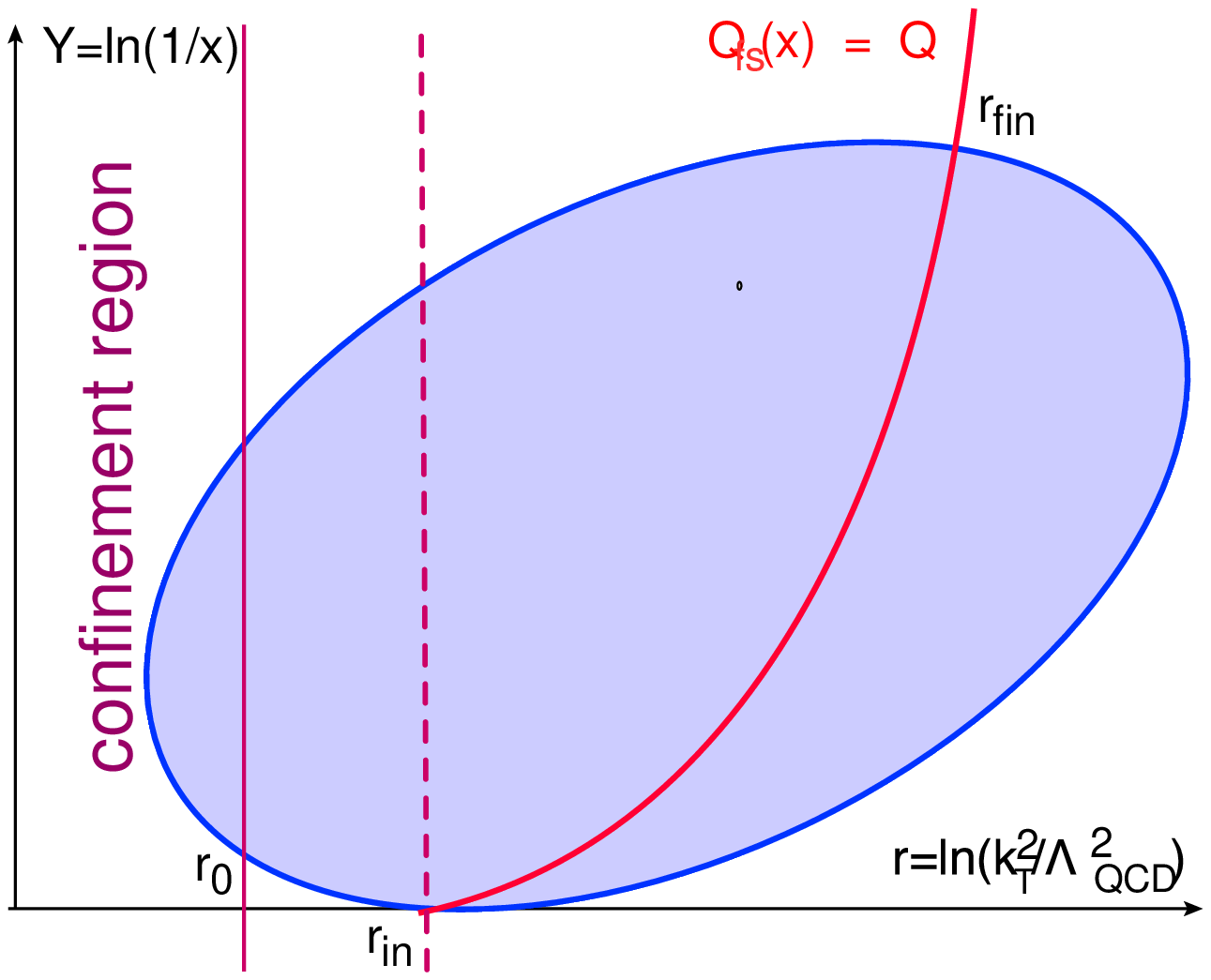}
      \caption{The distribution of the transverse momenta of gluon in the BFKL Pomeron for BFKL evolution from
      $q^2 = q^2_{in}$ ($ r =r_{in}$) which does not coincide with $r = r_0$ (Bartels sigar \cite{BARTS}). The solution to the equation $Q = Q_s\Lb x \Rb$ is shown in red. The region to the left of this curve is the saturation region of  the non linear evolution. The region to the right of the curve is the region where we can trust the linear BFKL equation.  The dashed red line shows the case when $r_0 = r_{in}$ which we consider in the paper. }
      \label{bs}
\end{figure}

\begin{figure}[ht]
 \begin{center}
  \leavevmode
      \includegraphics[width=10cm, height=6cm]{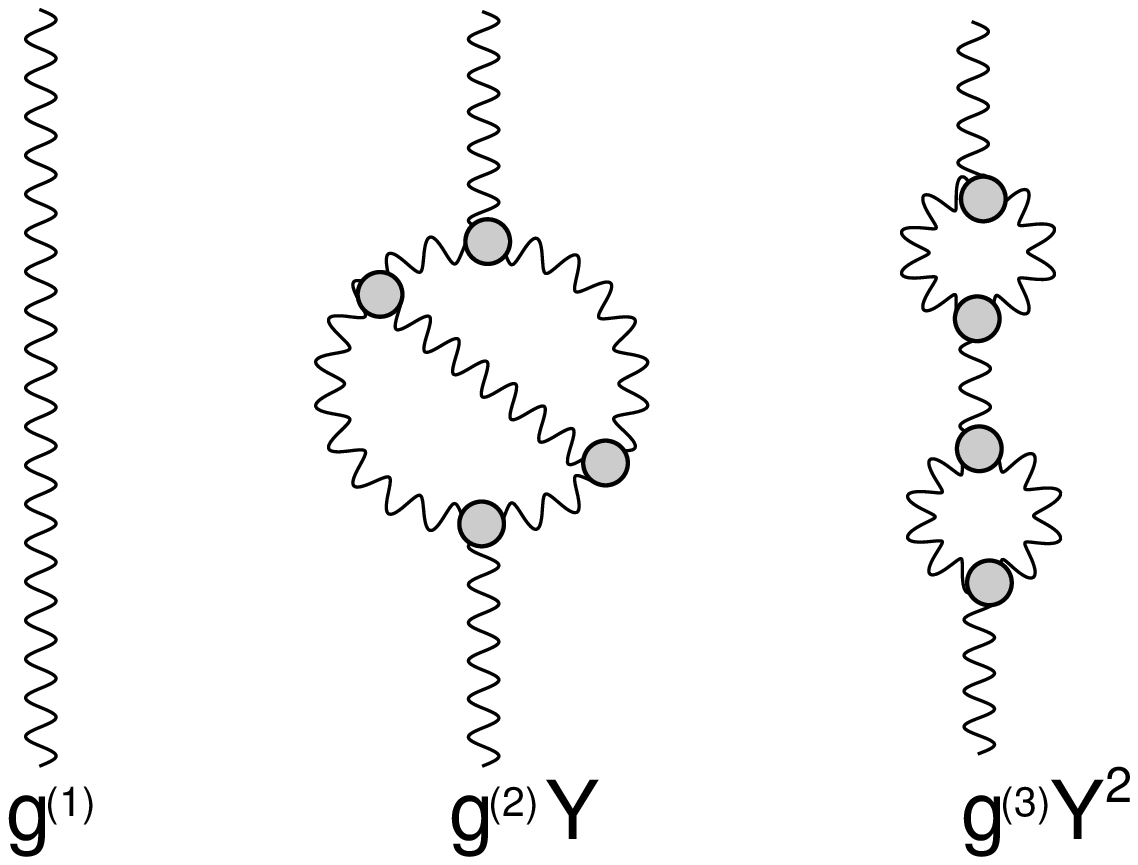}
      \end{center}
      \caption{The examples of the Pomeron diagrams that lead to $Y$ and $Y^2$ dependence in \protect\eq{GY}.  The wave lines denote soft Pomerons. }
      \label{pomen}
\end{figure}

Finally, the spectrum of the BFKL Pomeron depends only on initial value of $r_0= \ln\Lb k^2_{0,\bot}\Lambda^2_{QCD}\Rb$ while the residues depend on the measured $r$. All features of these poles are the same as in the procedure suggested in Refs. \cite{RevLI,KLR1,KLR2,KLRW}. The difference of our approach in comparison with the approach of those papers, is in the specific form how we impose the confinement on the BFKL equation.  It is well known that the BFKL approach cannot be implemented without introducing the restriction that stem from the confinement region\cite{BARTS}. In \fig{bs} the
typical distribution of the gluon momenta in the BFKL Pomeron is presented.
 For the values of the transverse momenta $q \leq q_0$ the unknown mechanism of 
 confinement of quark and gluons  plays the dominant role. We took the following approach to introduce the confinement to the BFKL evolution: we put the initial condition at $q_{in}=q_0$ ( $N_{in}$ in \eq{SOLR}) and consider  the BFKL evolution  only for the transverse momenta
 of partons ($k_\bot \geq q_0$),

         %%%%%%%%%%%%%%%%%%%%%%%%%%%%%%%%%%%%%%%%%%%%%%%%%%%%%
 %%%%%%%%%%%%%%%%%%%%%%%%%%%%%%%%%%%%%%%%%%%%%%%%%%%
This initial condition should be determined from the non-perturbative QCD. The high energy phenomenology\cite{SPH} as well as N=4  SYM \cite{SYM} lead to 
\beq \label{SFTIC}
N_{in}(Y_0, r=r_0)\,\,=\,\,g_\pom\Lb Y_0\Rb\,e^{\Delta_\pom Y_0}\,\,+\,\,g_\reg\Lb Y_0\Rb\,e^{\Delta_\reg Y_0}\eeq
where  $\Delta_\pom$ ($\Delta_\reg$) is the Pomeron (secondary Reggeon) intercept, respectively. The physical meaning of the two terms in  \eq{SFTIC} is clear in the high energy phenomenology
 based on the Reggeon approach. The first contribution describes the contribution of the soft Pomeron and its intercept will be a parameter of our fit. Function $g_\pom$ is the residue of the Pomeron contribution in which we include also the $\ln Y_0$ dependence which can stem from the Pomeron interactions. The second term in \eq{SFTIC} is responsible for the exchange of the secondary Reggeon. We fix the value of $\Delta_\reg = - 0.5$ in our fit. For 
 $g_\pom\Lb Y_0\Rb $ and $g_\reg\Lb Y_0\Rb$ we assume the simple form
 \beq \label{GY}
 g_\pom\Lb Y_0\Rb\,\,=\,\,g^{(1)}_\pom\,+\,g^{(2)}_\pom\,Y_0\,+\,g^{(3)}_\pom\,Y^2_0\,;\,\,\,\,\,\,\,\,
 g_\reg\Lb Y_0\Rb\,\,=\,\,g^{(1)}_\reg\,+\,g^{(2)}_\reg\,Y_0\,+\,g^{(3)}_\reg\,Y^2_0.
 \eeq
The polynomial in $Y_0$ reflects the enhanced diagrams for Pomeron  interaction shown in \fig{pomen}.
         %%%%%%%%%%%%%%%%%%%%%%%%%%%%%%%%%%%%%%%%%%%%%%%%%%%%%
 %%%%%%%%%%%%%%%%%%%%%%%%%%%%%%%%%%%%%%%%%%%%%%%%%%%

%%%%%%%%%%%%%%%%%%%%%%%%%%%%%%%%%%%%%%%%%%%%%%%%%%%%%%%%%%%%%%%

\subsection{Main formulae}

%%%%%%%%%%%%%%%%%%%%%%%%%%%%%%%%%%%%%%%%%%%%%%%%%%%%%%%%%%%%%%%
In $\om$-representation \eq{POLE1} and \eq{GY} can be written in the form
\beq \label{GOM}
g^{\lambda_i}\Lb \om\Rb\,\,=\,\,\frac{g^{(1)}_i}{\om\,-\,\lambda_I}\,\,+\,\,\frac{g^{(2)}_i}{\Lb \om\,-\,\lambda_i\Rb^2}
\,\,+\,\,\frac{2\,g^{(3)}_i}{\Lb \om\,-\,\lambda_i\Rb^3}
\eeq
where $\lambda_1\,=\,\Delta_\pom$ and $\lambda_{2}\,=\,\Delta_\reg$.

Using \eq{GF},\eq{SOLR},\eq{mellin1} and \eq{mellin2} we can re-write the dipole-target amplitude:
\bea \label{FING}
N\Lb Y, r \Rb\,&=&\, \sum_{i =1}^2 \Bigg\{g^{(\lambda_i)} R\Lb \om\,=\,\lambda_i; r, r_0\Rb\,+\,g^{(2)}_i\frac{d R\Lb \om\,=\,\lambda_i; r, r_0\Rb}{d \om}|_{\om = \lambda_i}\,+\,g^{(3)}_i\frac{d^2 R\Lb \om\,=\,\lambda_i; r, r_0\Rb}{d \om^2}|_{\om = \lambda_i}\Bigg\}\,e^{\lambda_i Y} \nn\\
 &+&  \sum_{i =1}^2 \sum_{n=1}^\infty\,\frac{g^{\lambda_i}\Lb \om_n\Rb}{\om_n\,-\,\lambda_i}\,e^{\om_n Y }\,\frac{Ai\Lb \Lb r - \frac{\chi_0}{b \om_n}\Rb\Lb \frac{b \om_n}{D_0}\Rb^{1/3}\Rb}{Ai'_{\om = \om_n}}
 \eea
  where
  \beq \label{R}
  R\Lb \om, r, r_0\Rb\,\,=\,\, \frac{Ai\Lb \Lb r - \frac{\chi_0}{b \om}\Rb\Lb \frac{b \om}{D_0}\Rb^{1/3}\Rb}{Ai\Lb \Lb r_0 - \frac{\chi_0}{b \om}\Rb\Lb \frac{b \om}{D_0}\Rb^{1/3}\Rb  }\,\,\mbox{and}\,\,\,Ai\Lb \Lb r_0 - \frac{\chi_0}{b \om}\Rb\Lb \frac{b \om}{D_0}\Rb^{1/3}\Rb \xrightarrow{\om \to \om_n}\,\,Ai'_{\om = \om_n}\,\Lb \om - \om_n\Rb  
    \eeq
  
  All above formulae have been written in the momentum representation. For calculating $F_2\Lb Q, Y\Rb$ it is more convenient to use the coordinate representation going from the dipole transverse momentum to the size of the dipole.
  Such a transformation it is easy to do in \eq{FING} by just replacing $r = \ln\Lb k^2_\perp/\Lambda^2_{QCD}\Rb\,\,\to\,\,
  r = \ln\Lb 1/\Lb x^2_\perp\Lambda^2_{QCD}\Rb\Rb $ where $x_\perp$ is the dipole size.
         %%%%%%%%%%%%%%%%%%%%%%%%%%%%%%%%%%%%%%%%%%%%%%%%%%%%%
 \begin{figure}
 \begin{center}
  \leavevmode
      \includegraphics[width=9cm, height=4cm]{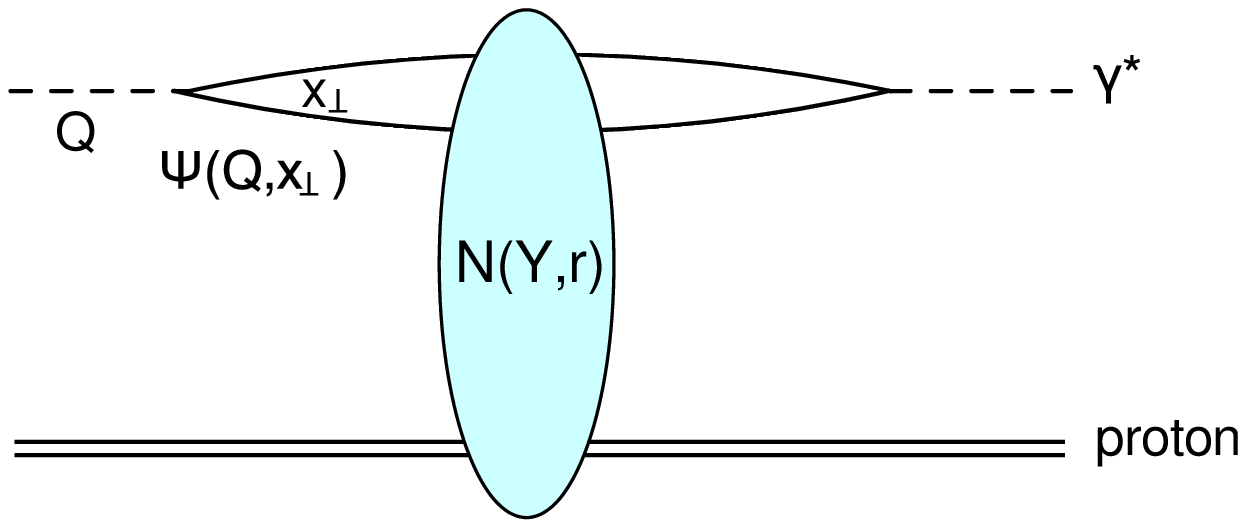}
      \end{center}
      \caption{ Deep inelastic scattering in dipole approach. Dashed line denotes a  photon with virtuality $Q$. The solid lines with arrows describe the quarks and antiquarks. $N\Lb Y, r = \ln\Lb 1/\Lb x^2_\perp \Lambda^2_{QCD}\Rb\Rb\Rb$ is the dipole-proton scattering amplitude, $Y = \ln \Lb 1/x_{Bj}\Rb$ where $x_{Bj} $ is the Bjorken $x$. }
      \label{dipr}
\end{figure} 
 %%%%%%%%%%%%%%%%%%%%%%%%%%%%%%%%%%%%%%%%%%%%%%%%%%%

  For calculating the amplitude for the deep inelastic scattering we need to recall that this process 
  happens through the virtual photon fluctuating into a $q \bar{q}$ pair(dipole) with the $q \bar{q}$ pair  
  proceeding to interact with the target \cite{GRIB,BJ,FS}. The cross section for the DIS process in this dipole picture can be written as follows\cite{KLZ,BER,MU90,NZ}
  \begin{align}\label{XS}
  \sigma_{tot}^{T,L} \Lb \gamma^\ast +\mbox{proton}| Y\,=\,\ln(1/x_{Bj}), Q^2\Rb\, = \, \int \frac{d^2 x_\perp}{4
    \, \pi} \, \int\limits_0^1 \, \frac{dz}{z \, (1-z)} \ 
  |\Psi^{\gamma^* \rightarrow q {\bar q}}_{T,L} ({\vec x}_\perp, z)|^2 \ 
  \sigma_{tot}^{q {\bar q} ,\mbox{proton}} ({\vec x}_\perp, Y).
\end{align}     

where $|\Psi_{T,L}|^2$ is the probability to find the dipole with size  $x_\perp$ into virtualphoton with transverse  or longitudinal polarization; and $\sigma_{tot}$ is the total cross section of $q \bar{q}$ (dipole) interaction with the proton.
The wave function of the virtual photon are known \cite{BJ,NZ}
\bea 
  |\Psi_{T}^{\gamma^* \rightarrow q {\bar q}} \Lb {\vec x}_\perp, z\Rb|^2
  \, &=& \, 2 \, N_c \, \sum_f \, \frac{\alpha_{em} \, Z_f^2}{\pi} \, z
  \, (1-z) \left\{ a_f^2 \, \left[ K_1 (x_\perp
      \, a_f) \right]^2 \, [z^2 + (1 - z)^2] + m_f^2 \, \left[ K_0
      (x_\perp \, a_f) \right]^2 \right\};\nn\\
 |\Psi_{L}^{\gamma^* \rightarrow q {\bar q}} \Lb{\vec x}_\perp, z\Rb|^2
  \, &=& \, 2 \, N_c \sum_f \frac{\alpha_{em} \, Z_f^2}{\pi} \, 4 \, Q^2
  \, z^3 \, (1 - z)^3 \, \left[ K_0 (x_\perp \, a_f) \right]^2.\label{PSILT}
  \eea
  where 
  \beq
  \label{AF}
  a^2_f\,\,=\,\,z (1 - z) Q^2\,\,+\,\,m^2_f,
  \eeq
  $\alpha_{em}$ is the fine-structure constant and $Z_f$ is the fraction of the electron charge that carries by the quark(antiquark) with flavour $f$ and mass $m_f$. 
  
  Finally, we need to recall that 
  \beq \label{F2}
 F_2 (x, Q^2) \, = \, \frac{Q^2}{4 \, \pi^2 \, \alpha_{em}} \,
  \sigma_{tot}\Lb \gamma^* + \mbox{proton}\Rb \, = \, \frac{Q^2}{4 \, \pi^2 \,
    \alpha_{em}} \, \left[ \sigma^{T}_{tot}\Lb \gamma^* +  \mbox{proton}\Rb + \sigma^{L}_{tot}\Lb \gamma^* +  \mbox{proton}\Rb \right] 
\eeq
\section{Description of the HERA data}
Using  formulae of the previous subsection we describe the HERA data on the deep   inelastic structure function $F_2$. This set of data was published in Ref.\cite{HERAPDF} and presents the combined data set of ZEUS and H1 collaborations.  The experimental errors are small and to describe these data is a challenge for any theoretical approach. In our procedure of the description we see two sets of the phenomenological parameters:  the intercept of the 
soft Pomeron $\lambda_1$ and two functions $g_\pom(Y_0)$ and $g_{reg}\Lb Y_0\Rb$, which are characterized the initial non-perturbative function of $x_{Bj}$ ( $Y = \ln(1/x_{Bj})$) at $Q^2=Q^2_0$ ($ r = r_0$); and two inputs for the $Q^2$ evolution: the initial val;ue of $Q=Q_0$ from which we start the evolution in $\ln(Q^2)$ ( $Q > Q_0$) and the mass of the quarks ($m_f$). It turns out that the value of $m_f$ in all our fits $\leq \,10 MeV$ and, therefore, we are dealing with current quarks as it should be in our approach.  

\begin{figure}
  \leavevmode
  \begin{tabular}{c c }
      \includegraphics[width=8cm]{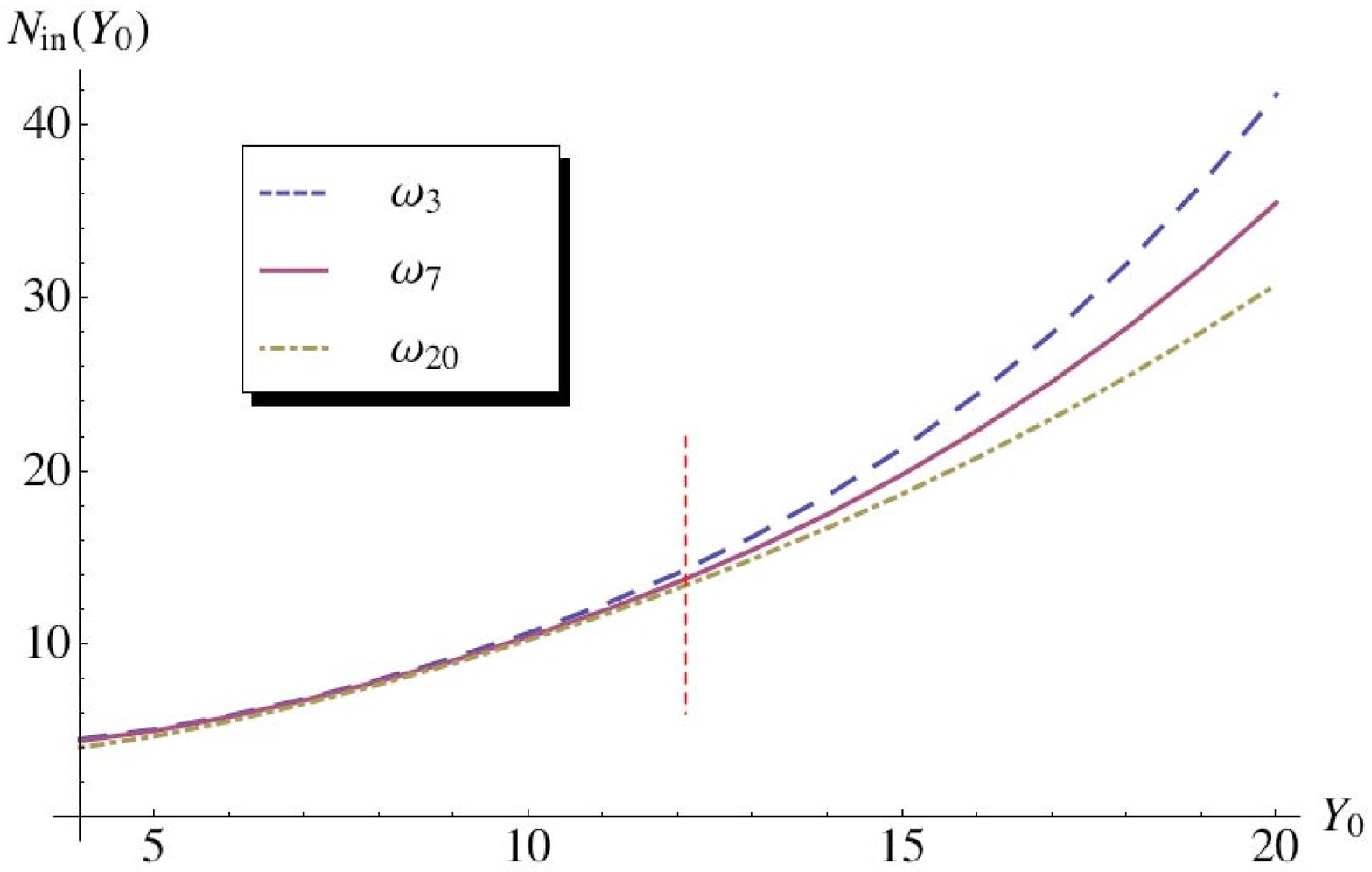}&      \includegraphics[width=8cm]{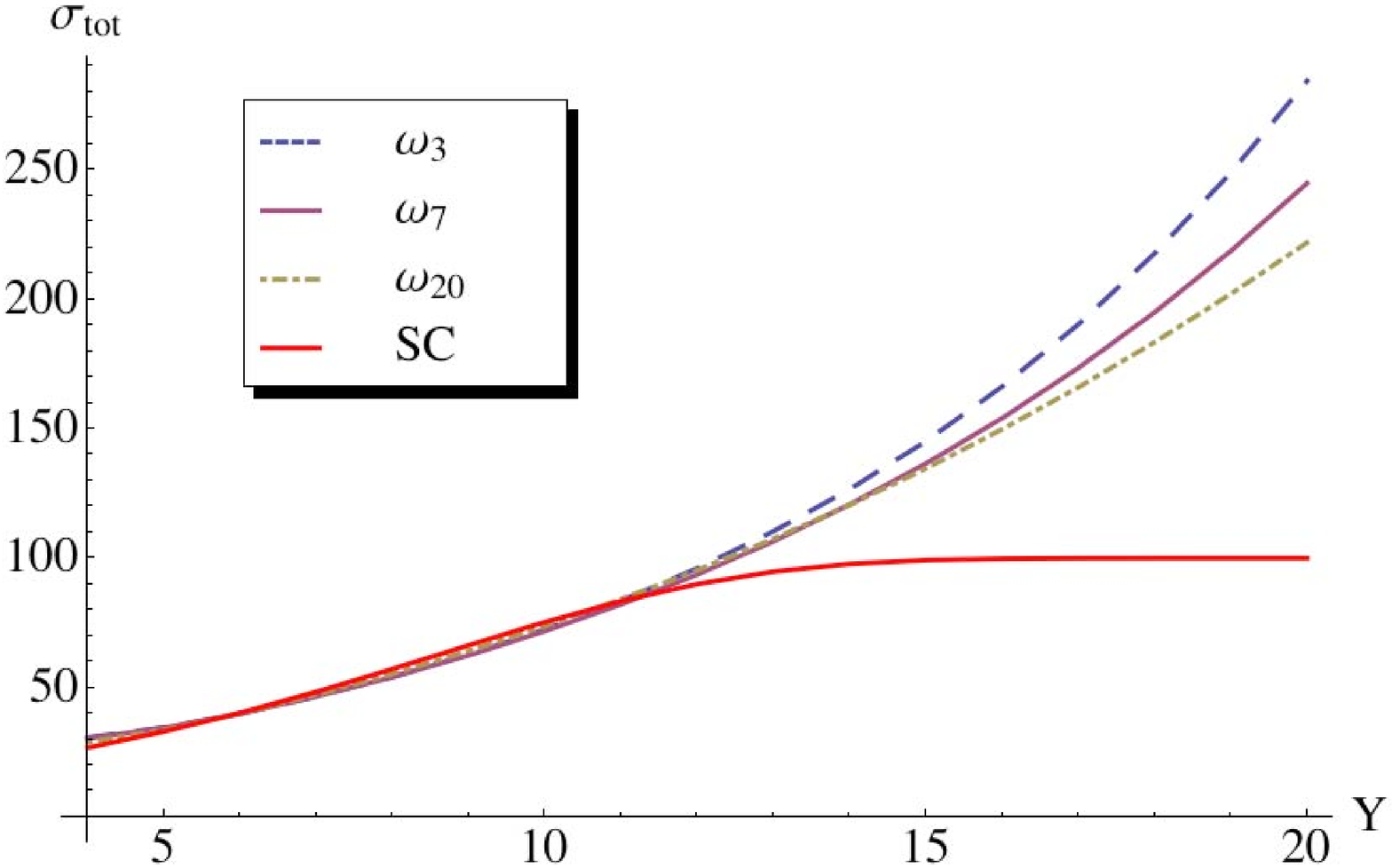}  \\
   \fig{incon}-a & \fig{incon}-b \\
   \end{tabular}
      \caption{\protect\fig{incon}-a: $N_{in}\Lb Y_0,r_0\Rb$ versus $Y_0$ for different solutions in our fit. $\lambda_1^{(n)}$ are shown in the legend. The vertical line shows the maximal value of $Y_0$ in  HERA experiment which we took into account in our fit. \protect\fig{incon}-b:  the same function $N_{in}\Lb Y_0,r_0\Rb$ as in \protect\fig{incon}-a but normalized to the value of the total cross section for proton-proton interaction at $W=20 \,GeV$. $Y = \ln\Lb s/s_0\Rb$ with $s_0 = 1 \,GeV^2$. The red solid curve gives \protect\eq{SCFIT} in the text with $\sigma_0 = 100 mb$, $\kappa = 0.115$ and $\Delta = 0.25$.   
      }
      \label{incon}
\end{figure}

As far as the fit of the initial function of $Y_0$, it turns out that we  have a set of fits with different values of the parameters(see Table 1). One can see from this table that we found the set of solutions which have in common the fact that $\lambda_1^{(n)}$ are close to the position of the poles in the Green function $\om_n$.  The differences between $\lambda_1^{(n)} - \om_n$ is so small
that parameter $\Lb \lambda_1^{(n)} - \om_n\Rb Y \ll 1 $ for the HERA kinematic region. It means that we actually claim that the intercept of the soft Pomeron coincides with the one of  poles that appears in the Green function. In this situation we need to rewrite \eq{FING} selecting separately the contribution with $\lambda_1^{(k)} =  \om_k$: viz
\bea \label{FING1}
N\Lb Y, r \Rb\,&=&\,  \Bigg\{g^{(1)}_1 Y\,+\,\h g^{(2)}_1 Y^2\,+\,\frac{1}{3} g^{(3)}_1 Y^3\Bigg\} \frac{Ai\Lb \Lb r - \frac{\chi_0}{b \om_k}\Rb\Lb \frac{b \om_k}{D_0}\Rb^{1/3}\Rb}{Ai'_{\om = \om_k}}e^{\om_k Y}\nn\\
&+& \, \Bigg\{g^{(\lambda_2)} R\Lb \om\,=\,\lambda_2; r, r_0\Rb\,+\,g^{(2)}_2\frac{d R\Lb \om\,=\,\lambda_2; r, r_0\Rb}{d \om}|_{\om = \lambda_2}\,+\,g^{(3)}_2\frac{d^2 R\Lb \om\,=\,\lambda_2; r, r_0\Rb}{d \om^2}|_{\om = \lambda_2}\Bigg\}\,e^{\lambda_2 Y} \nn\\
 &+&  \sum_{i =1}^2 {\sum_{n=1}^\infty}'\,\frac{g^{\lambda_i}\Lb \om_n\Rb}{\om_n\,-\,\lambda_i}\,e^{\om_n Y }\,\frac{Ai\Lb \Lb r - \frac{\chi_0}{b \om_n}\Rb\Lb \frac{b \om_n}{D_0}\Rb^{1/3}\Rb}{Ai'_{\om = \om_n}}
 \eea
where $\sum'$ denotes the sum without the term with $n = k$.
 It should be stressed that in spite of the fact that the largest contribution stems from one term in sum in \eq{FING}, we have to sum up to $n = N_{max} \approx 200$ to obtain the accuracy of our calculation smaller than the experimental errors.
All these solutions lead to good $\chi^2/d.o.f$ and the reason why we have them is clear from 
\fig{incon}-a in which we plotted the values of $N_{in}\Lb Y_0,r_0\Rb$ in \eq{SOLR}. One can see that in the HERA kinematic range ( to the left from the vertical line in \fig{incon}-a) all solutions give the same $N_{in}$ and the difference started to be visible only for larger values of $Y_0$.
 %%%%%%%%%%%%%%%%%%%%%%%%%%%%%%%%%%%%%%%%%%%%%%%%%%%%

%%%%%%%%%%%%%%%%%%%%%%%%%%%%%%%%%%%%%%%%%%%%%%%%%%%%
It should be stressed that our initial condition cannot be describe by the contribution of only two Regge poles: Pomeron and the secondary trajectory. We need to take into account the interaction of the Pomerons.  On the other hand in our parameterization we restrict ourselves by contribution of the enhanced diagrams (see \fig{pomen}), In other words it looks that we do not need to take into account the screening corrections. However this conclusion is premature since the simple formula with screening corrections:
\beq \label{SCFIT}
\sigma_{tot}\,\,=\,\,\sigma_0 \Big( 1\,-\,\exp\Lb - \kappa e^{\Delta Y}\Rb\Big)
\eeq
is able to describe the initial condition in the HERA kinematic region and leads to qualitatively reasonable values of the total cross sections at large $Y$ (see \fig{incon}-b)\footnote{We apply  our initial conditions  to  a description of proton-proton total cross sections as function of energy using the fact that they describe the enhanced diagrams.  In this way of doing we  chose the vertex of interaction of the soft Pomeon (see diagrams of  \fig{pomen}) from the conditiom that $\sigma_{tot}(\mbox{proton-proton})$ = 40  mb at $W =20 GeV$.}.   It worthwhile mentioning that $\Delta$ that gives the description,  is rather large  ($\Delta = 0.25$)   in agreement with the recent outcome from high energy Regge phenomenology \cite{SPH}.

Our main fitting parameter that is responsible for $Q^2$ evolution is $r_0$. We found that the best $\chi^2/d.o.f. $ we obtain for $q^2_0 = 0.25\,GeV^2$ ($r_0 = 1.83$) for any choice of $\lambda^{(k)}=\omega_k$.  However, the minimum of $\chi^2/d.o.f. $ is rather shallow.  The best $\chi^2/d.o.f. $ we found for  $\lambda^{(k)}= \omega_k = \omega_7$. (see Table 1).

\begin{figure}[t]
\vspace{3cm}
 \begin{tabular}{l}
  \leavevmode
      \includegraphics[width=12cm]{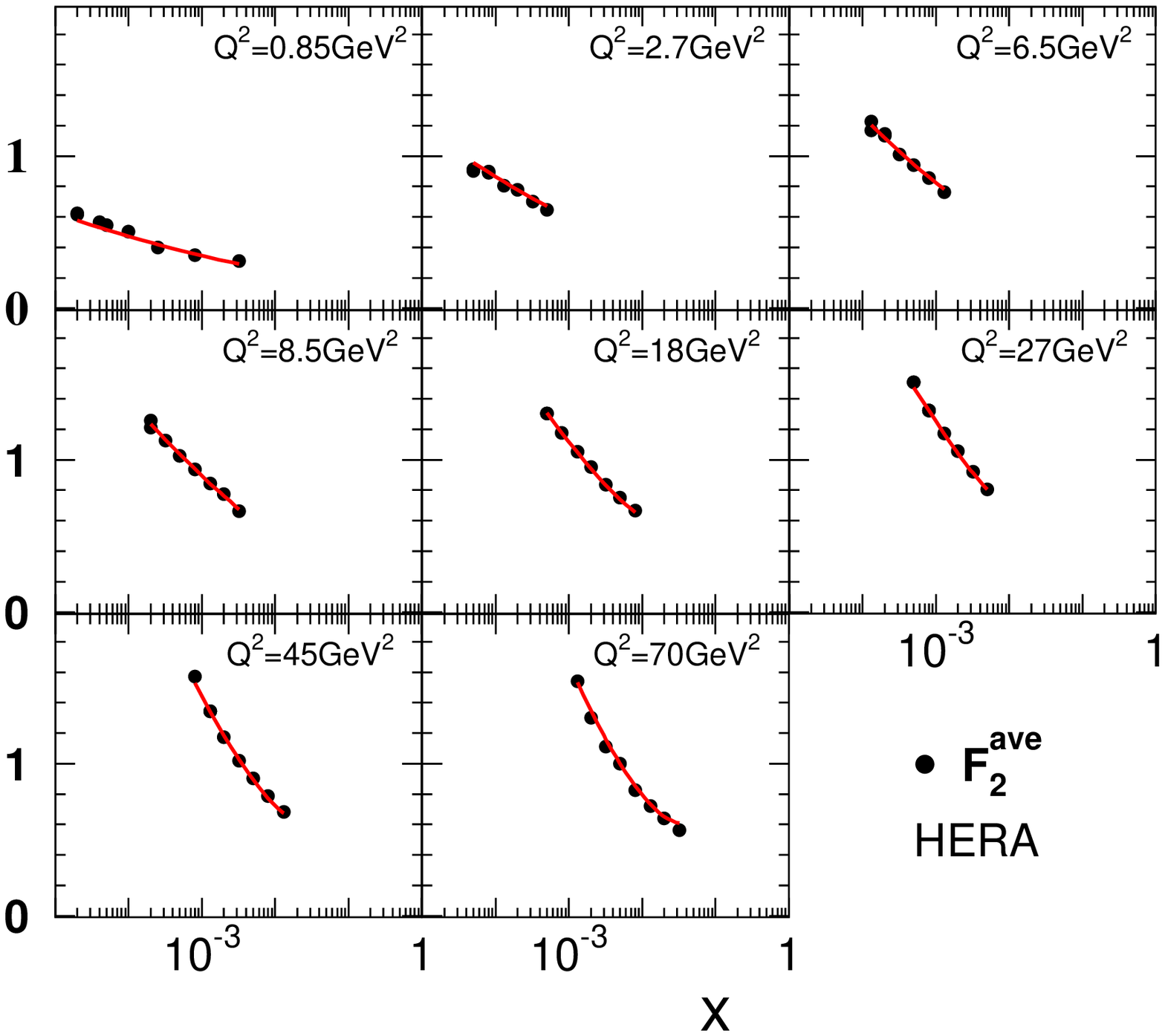}\\
      \end{tabular}
\vspace{3cm}
      \caption{ The deep inelastic structure function $F_2$ versus $x$. The data are taken from Ref.\cite{HERAPDF}. $r_0 = 1.83$. All other parameters in Table 2 for $\lambda^{(7)}_1$.}
      \label{fitx}
\end{figure}

\begin{figure}
 \begin{center}
  \leavevmode
      \includegraphics[width=10cm]{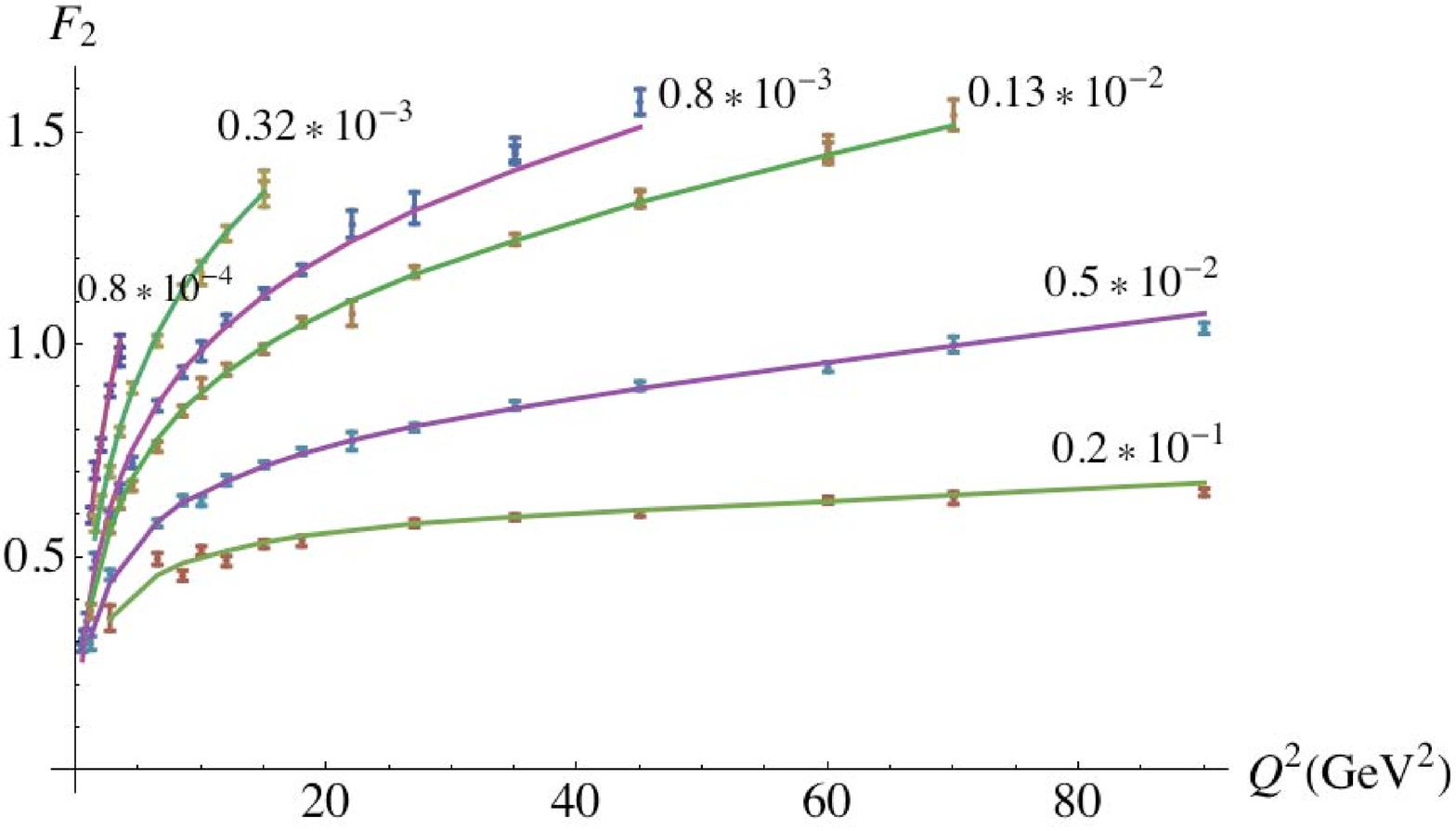}
      \end{center}
      \caption{ The deep inelastic structure function $F_2$ versus $Q$ at fixed $x_{Bj}$. The data are taken from Ref.\cite{HERAPDF}. $r_0 = 1.83$. All other parameters in Table 1 for $\lambda^{(7)}_1 = \om_7$. The values of $x_{Bj}$ are shown in the figure. }
      \label{fitq}
\end{figure}

The quality of the fit one can see from \fig{fitx} and \fig{fitq}

~

         %%%%%%%%%%%%%%%%%%%%%%%%%%%%%%%%%%%%%%%%%%%%%%%%%%%%%

 %%%%%%%%%%%%%%%%%%%%%%%%%%%%%%%%%%%%%%%%%%%%%%%%%%%

~

          %%%%%%%%%%%%%%%%%%%%%%%%%%%%%%%%%%%%%%%%%%%%%%%%%%%%%
 %%%%%%%%%%%%%%%%%%%%%%%%%%%%%%%%%%%%%%%%%%%%%%%%%%%

\begin{figure}[t]
~

~

~

~

~
  \leavevmode
  \begin{tabular}{c c c }
      \includegraphics[width=6cm,height=6cm]{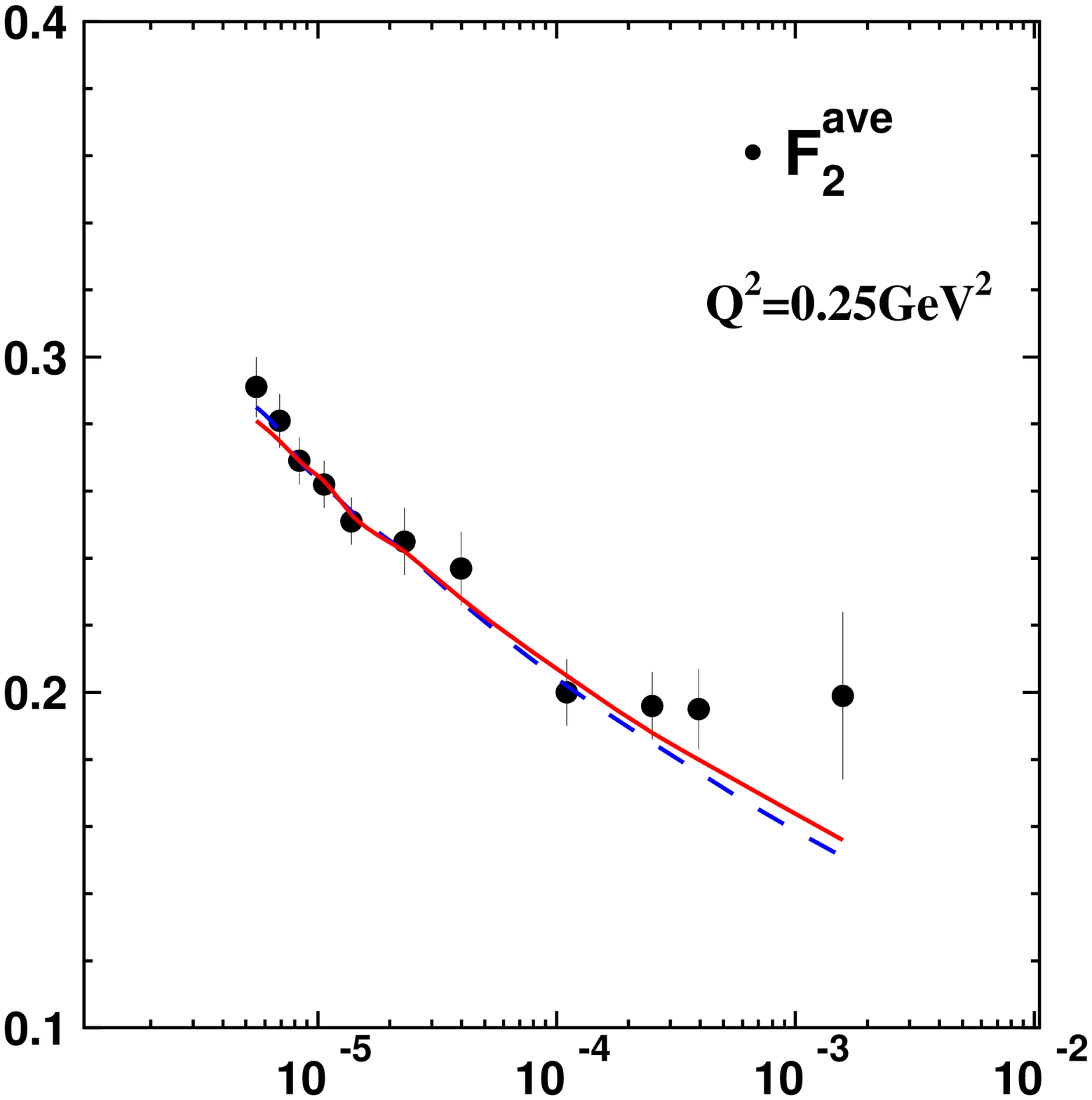}&  ~~~~~~~~~~~~~~~~~&    \includegraphics[width=6cm,height=6cm]{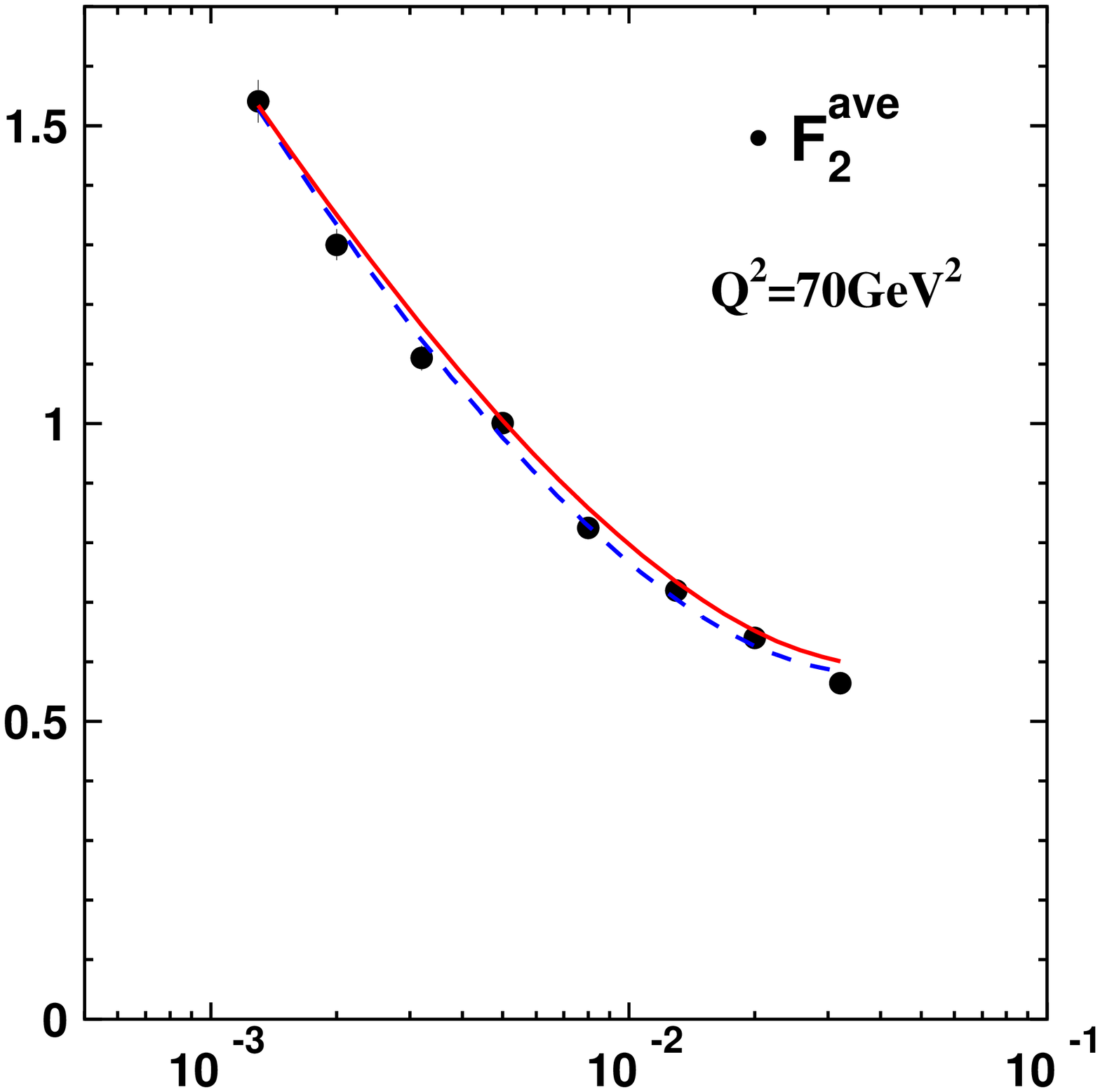}  \\
   \fig{compr}-a &~~~~~~~~& \fig{compr}-b \\
   \end{tabular}
      \caption{ Comparison of two solutions in our fit with $\lambda_1^{(3)}$ (dashed line) and $\lambda_1^{(7)}$ (solid line).
      }
      \label{compr}
\end{figure}

 Different solutions give the same descriptions: see \fig{compr} in which we compare the solution with $\lambda_1^{(3)}$ and  $\lambda_1^{(7)}$ .

\begin{figure}[h]
  \leavevmode
  \begin{center}
      \includegraphics[width=8cm]{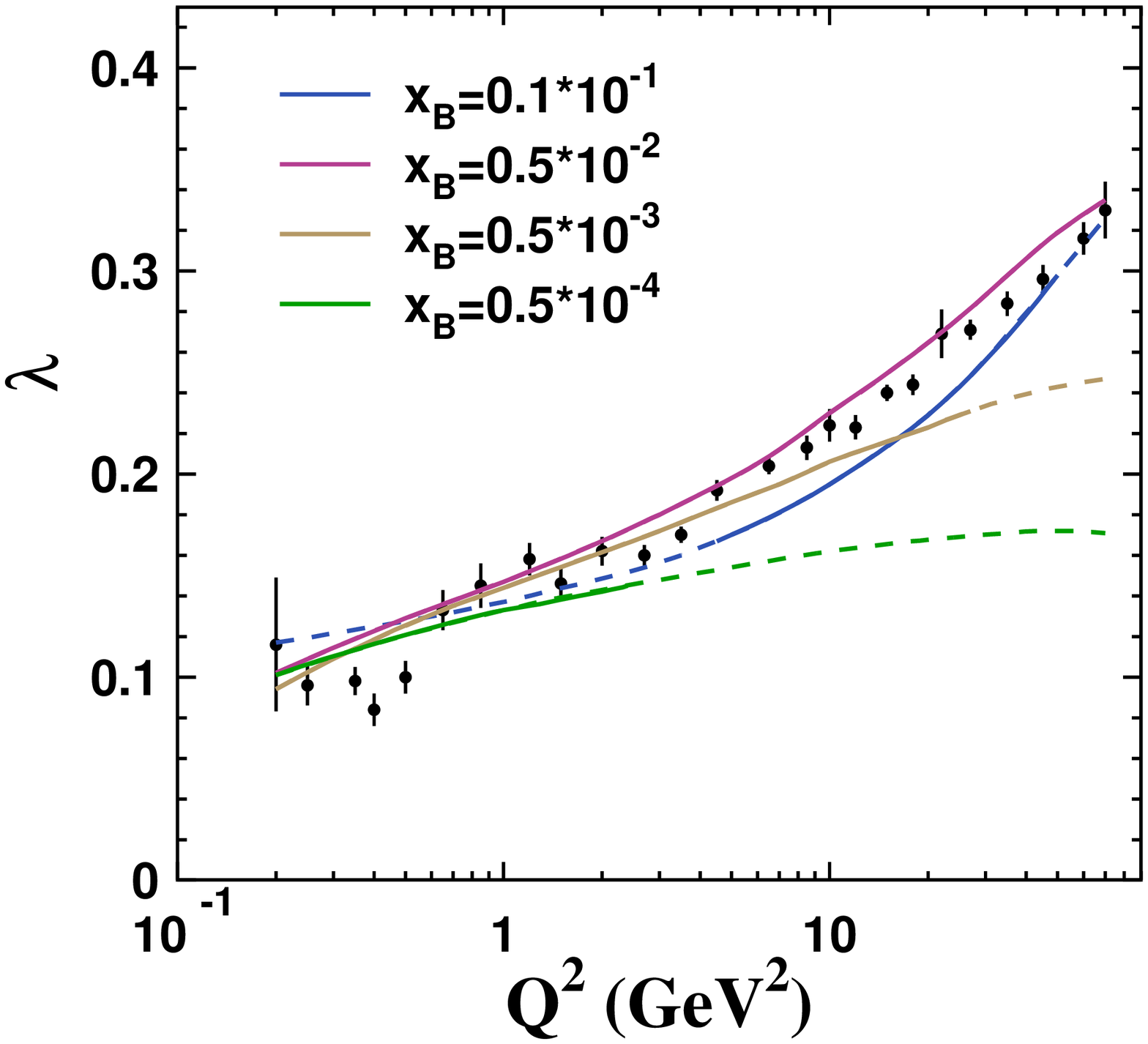}
   \end{center}
      \caption{ $d \ln F_2\Lb x_{Bj}, Q^q\Rb/d \ln(1/x_{Bj})$ versus $Q^2$ at different values of $x_{Bj}$ which are shown in the figure. The solid curves describe $d \ln F_2\Lb x_{Bj}, Q^q\Rb/d \ln(1/x_{Bj})$ in the kinematic region of HERA experiment while the dashed curve correspond to the kinematic region outside the HERA region and can be viewed as the predictions.  The data points shown in this figure were extracted from the experimental data of Ref.\cite{HERAPDF} by).   }
      \label{der}
\end{figure}

In \fig{der} we plot the calculated value of  $d \ln F_2\Lb x_{Bj}, Q^2\Rb/d \ln(1/x_{Bj})$ at different values of $x_{Bj}$. The solid lines corresponds to the  kinematic region in which we fit the data. The dashed curves can be considered as predictions.  One can see that we predict the dependence of this observable on $x_{Bj}$ but this dependence is rather mild in the HERA kinematic region.

    \section{Conclusions}
       In this paper we developed approach based on the BFKL evolution in $\ln\Lb Q^2\Rb$. We show that the simplest diffusion approximation with running QCD coupling is able to describe the HERA experimental data on the
       deep inelastic structure function with good $\chi^2/d.o.f. \approx 1.3$. We consider this result as the strong argument against the wide spread opinion that the BFKL dynamics has not been seen experimentally at HERA. 
  This result confirms the outcome of Refs. \cite{KLR1,KLR2,KLRW}, in which the BFKL equation was considered as  the theory of the reggeons.
  
  From our description of the  experimental data we learned several lessons:
  \begin{itemize}
  \item \quad The non-perturbative physics at long distances started to show up at $Q^2 = 0.25\,GeV^2$;
  
   \item \quad    The scattering amplitude at  $Q^2 = 0.25\,GeV^2$ cannot be written as sum of soft Pomeron and the secondary Reggeon but the Pomeron interactions should be taken into account;
   
      \item \quad    The Pomeron interactions can be reduced to the enhanced diagrams and, therefore, we do not see any needs for the shadowing corrections at HERA energies;
            \item \quad  We demonstrated that the shadowing correction could be sizable at higher than HERA energies without any contradiction with our initial conditions.
            \end{itemize}
            
    We believe that these lessons as well as the fact that we can reach a good description of the HERA data  in the framework of the BFKL dynamics, can be useful for future attempts to     understand the interface between long (soft)  and short(hard)  distance physics.    
            
   \section*{Acknowledgements}         
            We thank our colleagues at UTFSM and Tel Aviv university for encouraging discussions. Our special thanks goes to  Clara Salas  who shared with us the results of Refs.\cite{ SA1,SA2} before publication.
This research was supported by the Fondecyt (Chile) grants 1100648 and 1130549.
       \begin{table}[p]
        \centering
        \rotatebox{90}{
 \begin{minipage}{\textheight}{   \footnotesize
\begin{tabular}{|l|l|l|l|l|l|l|l|}
\hline \hline
n of $\om_n$ & 3 &4 & 5&6&7&10&20\\
\hline
$\om_n$& 0.111736 &0.083668& 0.066874& 0.055697& 0.0477217& 0.033382& 0.01667870\\
\hline
 $\lambda^{(n)}_1$& 0.112511&  0.084046 &             
                                 0.067082  &  0.055832&         
                               0.0478103 &  0.0333480&       
                               0.01668489\\ \hline
  $g^{(1)}_\pom$ &    2.520 $\pm$ 0.063&  2.622$\pm$ 0.088&  2.677 $\pm$ 0.058&
    2.639$\pm$ 0.083& 2.682 $\pm$ 0.081&  2.563$\pm$ 0.075&  2.349 $\pm$ 0.101 \\
    \hline
 $g^{(2)}_\pom$ &     0.099 $\pm$ 0.018&  0.141$\pm$ 0.015&  0.184 $\pm $ 0.013&
   0.203 $\pm$ 0.025& 0.221 $\pm$ 0.025&  0.116$\pm$ 0.024&  0.302 $\pm $ 0.032 \\
   \hline
$g^{(3)}_\pom$ &      0.000 $\pm$ 0.001&  0.006$\pm$  0.002&  0.009 $\pm$ 0.001&
 0.014  $\pm$  0.003& 0.016 $\pm$ 0.002&  0.027$\pm$  0.002&  0.034 $\pm$ 0.002 \\   \hline
$g^{(1)}_\reg$ &        8.999 $\pm$ 0.103&  8.260 $\pm$ 0.098&  8.784 $\pm$ 0.022& 
  7.756 $\pm$ 0.151 &  8.264 $\pm$ 0.152&  8.565 $\pm$ 0.132&  7.125 $\pm$ 0.188\\  
            \hline
  $g^{(2)}_\reg$ &    -2.448 $\pm$  0.036& -2.437 $\pm$ 0.089& -2.560 $\pm$ 0.016&
-2.432 $\pm$ 0.054& -2.518 $\pm$  0.054& -2.049 $\pm$ 0.048& -2.239 $\pm$ 0.071\\                   \hline
  $g^{(3)}_\reg $&     0.065  $\pm$ 0.003&  0.044 $\pm$  0.005&  0.035 $\pm$ 0.001&
   0.016$\pm$0.004&   0.012 $\pm$  0.004& -0.023 $\pm$ 0.004& -0.051 $\pm$ 0.006\\
      \hline
  $\chi^2/d.o.f$&349/227 = 1.54&310/227 = 1.36&  297/227 = 1.31 &299/227 = 1.32&      
 285/227 = 1.25&353/227 = 1.55& 368/227 = 1.62  \\
              \hline \hline
  \end{tabular}
\caption{The value of the fitted parameters for the initial condition in $Q^2$ evolution.}
}
\end{minipage}}
\end{table}

  %%%%%%%%%%%%%%%%%%%%%%%%%%%%%%%%%%%%%%%%%%%%%%%%%%%%

%%%%%%%%%%%%%%%%%%%%%%%%%%%%%%%%%%%%%%%%%%%%%%%%%%%%

  %%%%%%%%%%%%%%%%%%%%%%%%%%%%%%%%%%%%%%%%%%%%%%%%%%%%

%%%%%%%%%%%%%%%%%%%%%%%%%%%%%%%%%%%%%%%%%%%%%%%%%%%%

 \end{document}